\documentclass[journal=jacsat,manuscript=article]{achemso}

\usepackage[version=3]{mhchem} 
\usepackage{color}

\usepackage[normalem]{ulem}

\newcommand{\pg}[1]{\textcolor{black}{#1}}

\author{Akta Singh}
\affiliation{Tata Institute of Fundamental Research Hyderabad, Telangana 500046, India}
\author{Nayana Mukherjee}
\affiliation{Department of Mathematics. Mahindra University, Hyderabad, Telangana 500043, India}
\author{Jagannath Mondal}\email{jmondal@tifrh.res.in}
\affiliation{Tata Institute of Fundamental Research Hyderabad, Telangana 500046, India}
\author{Pushpita Ghosh}\email{pushpita@iisertvm.ac.in}
\affiliation{School of Chemistry, Indian Institute of Science Education and Research, Thiruvananthapuram, Kerala 695551, India}
\alsoaffiliation{Center for High-Performance Computing, Indian Institute of Science Education and Research, Thiruvananthapuram, Kerala 695551, India}

\title[An \textsf{achemso} demo]
  {
  Reaction--transport coupling drives spatiotemporal organization in fuel-driven supramolecular polymerization}

\abbreviations{IR,NMR,UV}
\keywords{American Chemical Society, \LaTeX}

\begin{document}


\begin{abstract}
\pg{Chemically fueled supramolecular systems provide a versatile platform for generating nonequilibrium structures and dynamical instabilities, including chemical oscillations and traveling waves reminiscent of biological organization. However, a minimal mechanistic framework capable of capturing the emergence of such spatiotemporal order is still lacking.} Here, we develop a minimal reaction--transport framework for fuel-driven supramolecular polymerization that couples activation--deactivation chemistry with cooperative assembly, fragmentation, and polymer length-dependent diffusion. The model captures autonomous oscillations arising through a Hopf bifurcation and demonstrates how temporal instabilities evolve into spatial self-organization upon inclusion of transport.
We show that the nonlinear interplay between reaction kinetics and state-dependent mobility gives rise to traveling polymerization fronts, oscillatory wave dynamics, and complex spatiotemporal patterns. The propagating fronts exhibit near-ballistic dynamics, revealing a fundamentally nonequilibrium transport mechanism emerging from reactive feedback and dynamically evolving diffusivity. These findings establish a minimal physical framework connecting dissipative self-assembly, nonlinear transport, and active matter, while providing design principles for programmable supramolecular materials capable of autonomous spatiotemporal organization.
  
\end{abstract}

\section{Introduction}
Self-organization and pattern formation are hallmarks of systems maintained far from thermodynamic equilibrium~\cite{cross1993pattern,turing1990chemical}. Such processes are ubiquitous in nature, from morphogenesis to intracellular organization, and arise from nonlinear coupling between reaction kinetics, energy input, and spatial transport~\cite{epstein1998introduction,murray2003spatial,cross2009pattern,karsenti2008self,halatek2018rethinking}. Living cells exemplify this principle: through continuous fuel consumption, they sustain nonequilibrium steady states (NESS) that drive complex behaviors such as oscillations, waves, and spatial organization~\cite{alberts2022molecular,hess2017non,huang2003dynamic}. Cytoskeletal filaments such as actin and microtubules undergo ATP- or GTP-fueled polymerization--depolymerization cycles that enable dynamic reorganization and spatiotemporal patterning~\cite{bugyi2010control,dominguez2011actin,brouhard2018microtubule,akhmanova2022mechanisms,mandelkow1989spatial,mondal2025role}. Inspired by these processes, chemically fueled supramolecular systems have emerged as a powerful platform for realizing life-like dynamics in synthetic materials~\cite{SoumenDe,dhiman2018temporally,howlett2022autonomously,ragazzon2018energy,leira2018oscillations,Peter2022,hou2023supramolecular,lang2023mechanosensitive,li2023programmable}. Despite these advances, it remains unclear what minimal physical ingredients are sufficient to generate such spatiotemporal organization from molecular-scale processes. In particular, how fuel-driven activation--deactivation cycles couple with cooperative assembly and transport to produce autonomous dynamics without external control is still not well understood.

Recent experiments have demonstrated that chemically fueled supramolecular systems can exhibit rich collective dynamics, including autonomous oscillations, propagating fronts, and centimeter-scale spatial patterns~\cite{leira2018oscillations,kubota2020force,te2018spatiotemporal,TenaSolsona2017,DDas2019}. In these systems, monomers are transiently activated by chemical or redox fuels to assemble into higher-order structures and subsequently deactivated, leading to dissipative self-assembly sustained by continuous turnover of components~\cite{leira2018oscillations,boekhoven2015transient,ragazzon2018energy,heuser2015generic}. These fuel-driven reaction cycles embody the essence of \emph{dissipative self-assembly}, where structural order is maintained only by ongoing energy dissipation~\cite{astumian2018kinetic,merindol,van2022out}. A key feature of such systems is the interplay between autocatalytic growth, fragmentation-mediated feedback, and inhibitory decay processes, which together can generate nonequilibrium steady states and dynamic instabilities. In parallel, theoretical studies in systems chemistry and soft active matter have established that coupling nonlinear reaction networks with spatial transport can give rise to oscillations, waves, and pattern formation~\cite{hess2017non,merindol,ragazzon2018energy,sagues2003nonlinear}. However, a unified minimal framework that directly connects molecular-scale supramolecular polymerization kinetics with emergent spatiotemporal organization remains lacking.

Here, we develop a minimal reaction--diffusion model of fuel-driven supramolecular polymerization that captures the essential feedback mechanisms underlying oscillations and spatial pattern formation. The model integrates monomer activation--deactivation cycles with cooperative polymer growth and fragmentation, sustained by a constant fuel flux. Linear stability analysis reveals the onset of temporal oscillations via a Hopf bifurcation, while coupling to diffusion generates traveling polymerization fronts and complex spatiotemporal patterns. Furthermore, we uncover accelerated wavefront propagation exhibiting super-diffusive scaling, establishing a direct link between molecular-scale reaction kinetics and emergent transport behavior. This framework provides a minimal mechanistic basis for self-organized dynamics in dissipative supramolecular systems and offers design principles for programmable, fuel-driven materials.

\section{Model Description}
\begin{figure*}[h!]
    \includegraphics[width=0.99\textwidth]{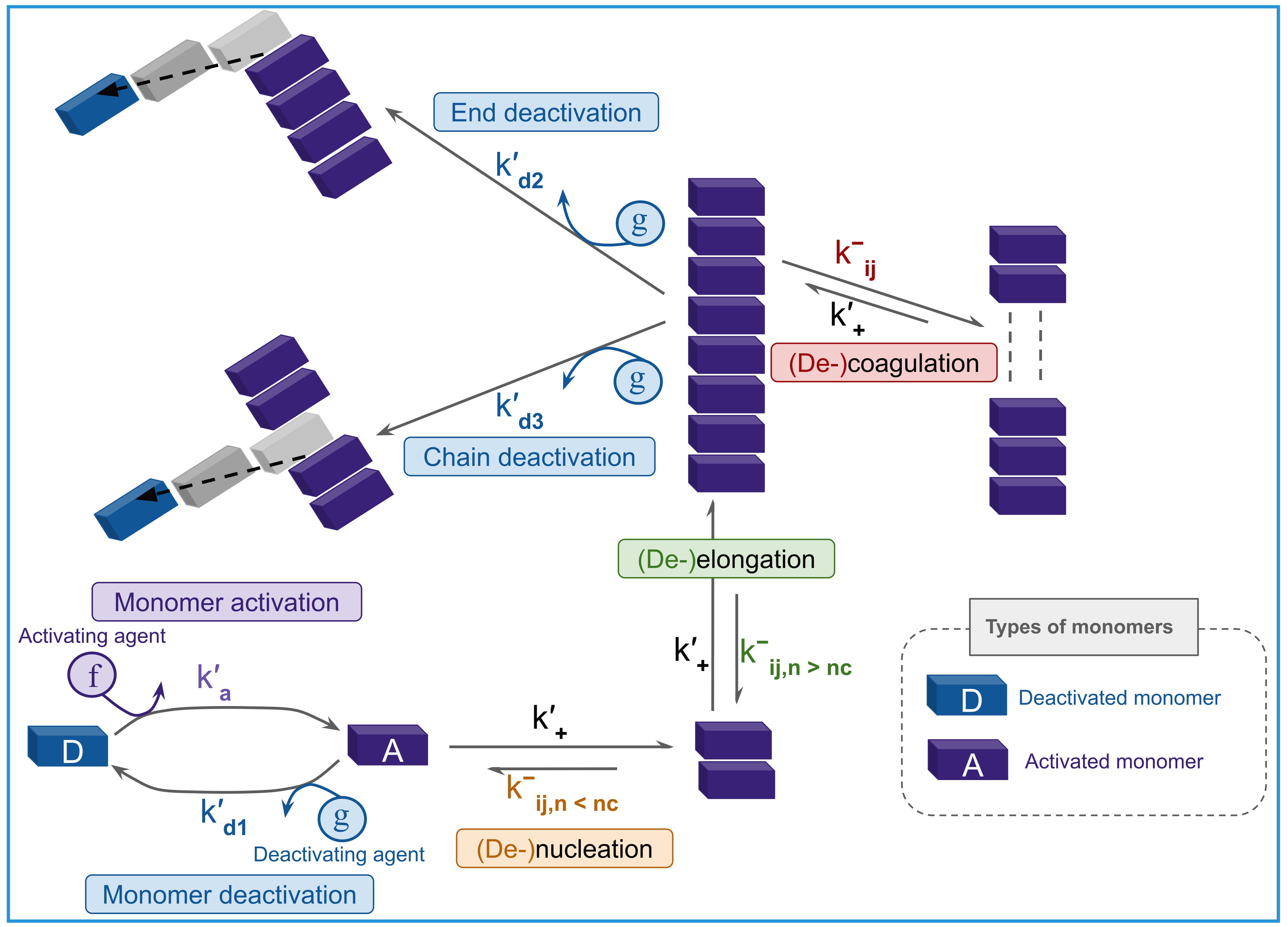}
    \caption{Schematic of the coupled cooperative supramolecular polymerization model~\cite{sharko2022insights}. Constant supplies of activating (\( f \)) and deactivating (\( g \)) agents drive monomer activation/deactivation, polymer assembly, and disassembly under non-equilibrium conditions.}
    \label{fig1}
\end{figure*}

To establish a baseline for the spatiotemporal dynamics discussed below, we first consider the temporal behavior of the spatially homogeneous (well-mixed) system. We build on a coupled cooperative supramolecular polymerization model introduced by Sharko \textit{et al.}~\cite{sharko2022insights}, schematically illustrated in Fig.~\ref{fig1}. In this framework, monomers undergo reversible activation and deactivation in a chemically fueled environment, while polymerization proceeds via nucleation, elongation, and coagulation, together with their reverse processes. Constant concentrations of activating and deactivating agents maintain the system in a nonequilibrium steady state.

A key feature of the model is the presence of competing feedback mechanisms. Polymer fragmentation increases the number of chain ends and thereby enhances growth (positive feedback), whereas monomer and polymer deactivation act as inhibitory processes that limit assembly (negative feedback). The interplay of these processes provides the basis for oscillatory and nonequilibrium dynamics.

The system dynamics are described in terms of four variables: the concentration of deactivated monomers \( d(t) \), activated monomers \( a_{1}(t) \), the total polymer number concentration \( m_{0}^\prime(t) = \sum_{n=2}^{\infty} a_{n}(t) \), and the total polymerized mass \( m_{1}^\prime(t) = \sum_{n=2}^{\infty} n a_{n}(t) \). The total monomer concentration is conserved, such that \( c_{\mathrm{tot}} = d(t) + a_{1}(t) + m_{1}^\prime(t) \).

To obtain a tractable description, we employ a reduced moment-based model that retains only the dominant kinetic processes governing the evolution of these variables in the oscillatory regime~\cite{sharko2022insights}. The reduced equations depend on the activation rate \( k_{a}^\prime \), the assembly rate \( k_{+} \), and the effective deactivation rates \( k_{d2}^\prime \) and \( k_{d3}^\prime \). A detailed derivation of the reduced model, including the moment-closure approximation and underlying assumptions, is provided in the Supporting Information (SI).

The resulting dimensionless equations are given by:
\begin{align}
\dot{d} &= -k_{a}^\prime d + 2 k_{d2}^\prime m_{0}^\prime,
\label{eq1}
\\
\dot{a_{1}} &= k_{a}^\prime d - 2 k_{+} a_{1} m_{0}^\prime,
\label{eq2}
\\
\dot{m_{1}^\prime} &= -2 k_{d2}^\prime m_{0}^\prime + 2 k_{+} a_{1} m_{0}^\prime,
\label{eq3}
\\
\dot{m_{0}^\prime} &= k_{d3}^\prime (m_{1}^\prime - 4 m_{0}^\prime)
- \frac{2 k_{d2}^\prime C m_{0}^\prime}{(m_{1}^\prime/m_{0}^\prime - 2)^{2}}
- 2 k_{d3}^\prime m_{0}^\prime
- k_{+} (m_{0}^\prime)^{2}.
\label{eq4}
\end{align}

Here, the parameter \( C \) captures the effective contribution of short polymer species within the moment-closure approximation. The model is nondimensionalized by setting \( k_{+} = K = 1 \), with concentrations scaled by \( K \) and time by \( (k_{+} K)^{-1} \). In the parameter regime considered here, the mean polymer length remains well above the nucleus size, ensuring the validity of the reduced description.

\section{Results and Discussion}
\subsection{Identifying the conditions for temporal oscillations}
\begin{figure*}[h!]
     \includegraphics[width=0.95\textwidth]{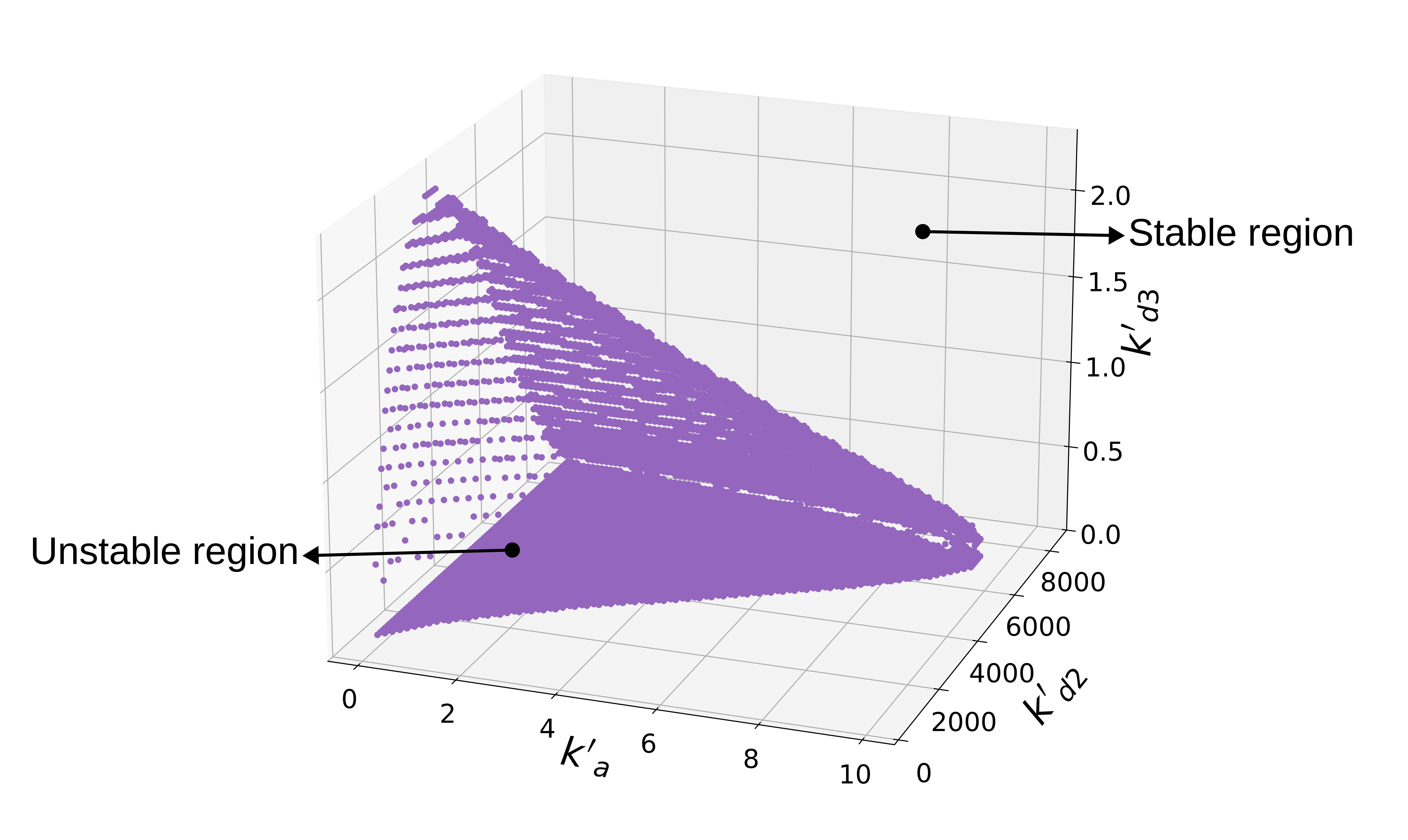}
   \caption{
Bifurcation diagram in the three-dimensional parameter space defined by 
$k_{a}^\prime$, $k_{d2}^\prime$, and $k_{d3}^\prime$, obtained from linear stability analysis of the 3-component reduced system (SI, Section-1). The bifurcation surface, shown as purple dots, separates the parameter space into linearly stable and unstable regions.
}\label{fig2}
\end{figure*}

We first analyze the temporal behavior of the well-mixed system in the absence of diffusion. The reduced model exhibits self-sustained oscillations over a broad range of parameters, arising from the interplay between autocatalytic growth and inhibitory deactivation processes. To identify the onset of oscillations, we perform a linear stability analysis of the homogeneous steady state (see Methods in SI). The resulting bifurcation diagram in the parameter space defined by $k_{a}^\prime$, $k_{d2}^\prime$, and $k_{d3}^\prime$ is shown in Fig.~\ref{fig2}. The bifurcation surface separates regimes of stable steady states from those exhibiting sustained oscillations. Crossing this boundary leads to a Hopf bifurcation, beyond which the system evolves into stable limit-cycle oscillations characterized by periodic growth and decay of polymer mass. The emergence of oscillations is governed by the balance between chain deactivation-induced positive feedback, which promotes growth, and delayed end deactivation processes, which provide negative feedback.

These results establish the minimal kinetic conditions required for autonomous temporal oscillations in fuel-driven supramolecular polymerization.

\subsection{Kinetic control of polymer length and dynamics}

\begin{figure*}[h!]
\includegraphics[width=\textwidth]{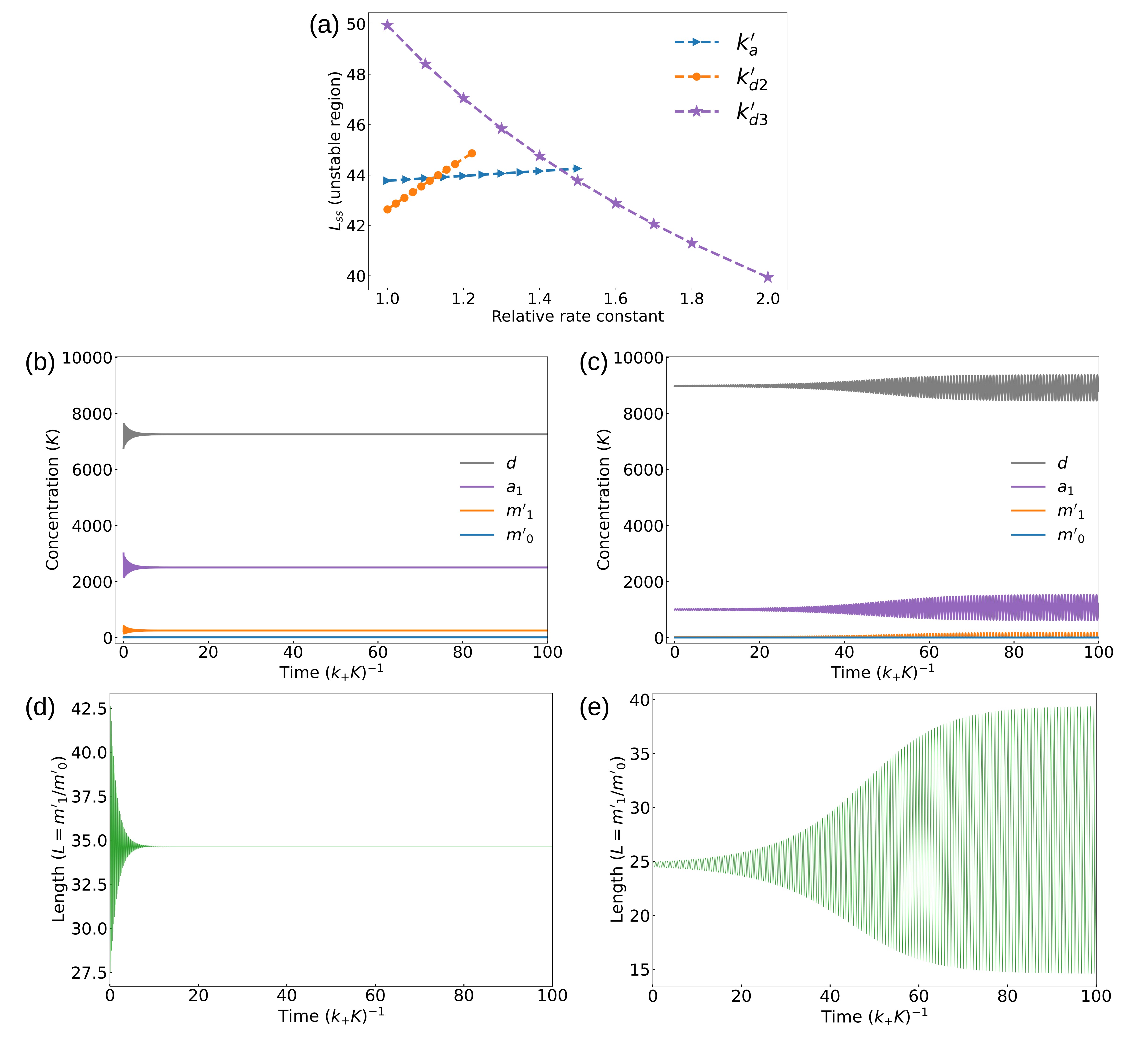}
\caption{
(a) Steady-state average polymer length $L_{ss} = m_{1}^*/m_{0}^*$ as a function of kinetic parameters in the oscillatory regime. 
(b, c) Temporal evolution of $d(t)$, $a_{1}(t)$, $m_{1}^\prime(t)$, and $m_{0}^\prime(t)$ in the stable and oscillatory regimes, showing damped relaxation and sustained oscillations, respectively. 
(d, e) Time evolution of the mean polymer length $L = m_{1}^\prime/m_{0}^\prime$ corresponding to (b) and (c).
}
\label{fig3}
\end{figure*}

We next examine how the key kinetic parameters - monomer activation ($k_{a}^\prime$), polymer-end deactivation ($k_{d2}^\prime$), and chain deactivation ($k_{d3}^\prime$) - control the steady-state polymer length and system dynamics. The average polymer length in the oscillatory regime is defined as $L_{ss} = m_{1}^*/m_{0}^*$ (Fig.~\ref{fig3}a).

Increasing the activation rate $k_{a}^\prime$ promotes the formation of longer polymers by enhancing the supply of activated monomers for growth. In contrast, polymer-end deactivation ($k_{d2}^\prime$) reduces both polymer number and mass, but more strongly suppresses the number of chains, leading to an increase in the average polymer length. Chain deactivation ($k_{d3}^\prime$), on the other hand, fragments longer polymers into shorter ones, increasing polymer number while reducing overall mass, and thus decreases the average length.

To characterize the dynamical response, we numerically integrate the reduced kinetic equations (see SI). In the stable regime, the system exhibits damped oscillations that relax to a steady state (Fig.~\ref{fig3}b), whereas in the oscillatory regime, the dynamics evolve into sustained limit cycles (Fig.~\ref{fig3}c). These results are consistent with the Hopf bifurcation identified above and confirm that the balance between growth and deactivation processes governs the emergence of temporal oscillations. The evolution of the mean polymer length $L = m_{1}^\prime/m_{0}^\prime$ (Fig.~\ref{fig3}d,e ) shows that $L$ remains well above the nucleus size throughout the dynamics, confirming the validity of the reduced moment-based description (detail in  SI). 

The oscillatory dynamics are robust over a range of parameter values of $C$ (see Figure~S1, in SI), indicating that the underlying feedback mechanisms provide a stable basis for nonequilibrium behavior. This kinetic instability provides the basis for the emergence of spatiotemporal structures upon inclusion of diffusion.
\begin{figure*}[h!]
\includegraphics[width=1\textwidth]{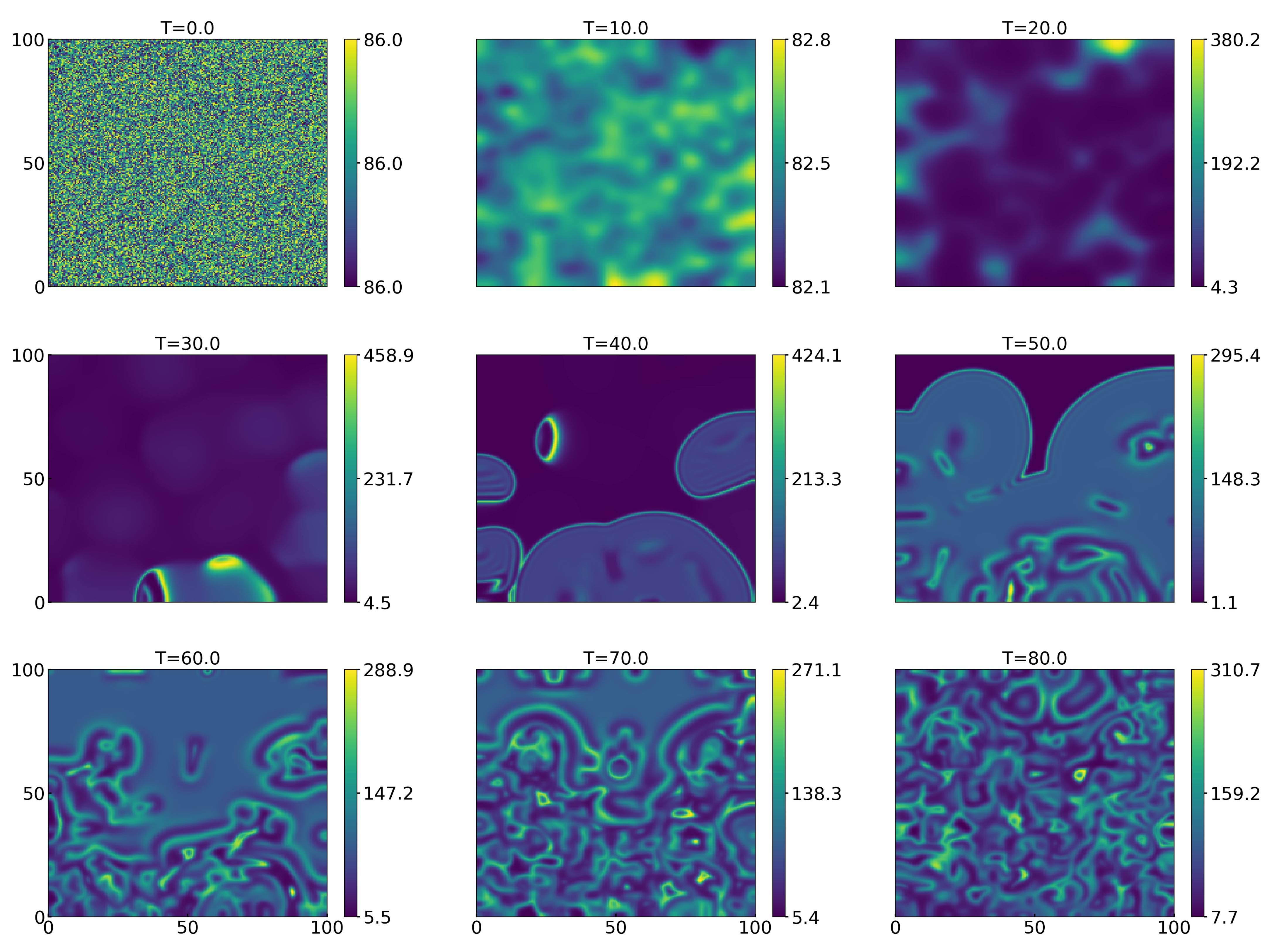}
\caption{
Snapshots of the polymer mass concentration at different times during the 
spatiotemporal simulation for a parameter set located in the oscillatory region 
($k_{a}^\prime = 4.0$, $k_{d2}^\prime = 5000.0$, and $k_{d3}^\prime = 1.5$), initialized with random perturbations. Domain size is $100 \times 100$. Grid size is 0.5. Time is expressed in units of $(k_{+} K)^{-1}$.}
\label{fig4}
\end{figure*}

\subsection{Emergence of self-organized wavefronts and patterns}
To explore spatiotemporal instability leading to self-organized pattern formation in the supramolecular polymerization system, we extend the kinetic model (Eqs.~\ref{eq1}--\ref{eq4}) by incorporating diffusion. Polymer transport is described using Rouse-type scaling~\cite{rouse1953theory,doi1986theory,rubinstein2003polymer}, in which the diffusion coefficient decreases inversely with polymer length, providing a physically meaningful description for dilute, unentangled supramolecular polymers. The resulting reaction--diffusion equations for the four species—deactivated monomers ($d$), activated monomers ($a_{1}$), total polymer mass ($m_{1}^\prime$), and total polymer number ($m_{0}^\prime$), are given by:
\begin{equation}
\dot{d} = -k_{a}^\prime d + 2k_{d2}^\prime m_{0}^\prime + D_{0}\nabla^{2}d,
\label{eq8}
\end{equation}
\begin{equation}
\dot{a_{1}} = k_{a}^\prime d - 2k_{+}a_{1}m_{0}^\prime + D_{0}\nabla^{2}a_{1},
\label{eq9}
\end{equation}
\begin{equation}
\dot{m_{1}^\prime} = -2k_{d2}^\prime m_{0}^\prime + 2k_{+}a_{1}m_{0}^\prime + {\vec{\nabla}\cdot (D_{m}\vec{\nabla} m_{1}^\prime)},
\label{eq10}
\end{equation}
\begin{equation}
\dot{m_{0}^\prime} = k_{d3}^\prime (m_{1}^\prime - 4m_{0}^\prime)
-\frac{2k_{d2}^\prime C m_{0}^\prime}{(m_{1}^\prime/m_{0}^\prime - 2)^{2}}
-2k_{d3}^\prime m_{0}^\prime - k_{+}(m_{0}^\prime)^{2}
+ {\vec{\nabla}\cdot(D_{m}\vec{\nabla} m_{0}^\prime)}.
\label{eq11}
\end{equation}
Here, $D_{0}$ and $D_{m}$ denote the diffusion coefficients of monomers and polymers, respectively. The monomer diffusion coefficient is kept fixed at $D_0=1.0$ in the dimensionless units. Following Rouse scaling, the polymer diffusivity depends on the instantaneous mean polymer length, $D_{m} = D_{0}/L_{p}$ with $L_{p} = m_{1}^\prime/m_{0}^\prime$, while monomer diffusion remains constant. This length-dependent transport introduces a direct coupling between polymerization kinetics and mobility: polymer growth locally suppresses diffusion, whereas depolymerization enhances it. This feedback provides a minimal physical mechanism linking molecular-scale assembly to emergent spatial organization.

We numerically integrate the reaction--diffusion equations for parameter values within the oscillatory regime, starting from a nearly homogeneous state with small random perturbations (see Supporting Information~(SI)). Initially, the system exhibits weak spatial fluctuations, which grow over time into localized regions of high polymer concentration. These regions act as nucleation centers that propagate outward as traveling wavefronts. As the dynamics evolve, the wavefronts expand, interact, and merge, giving rise to complex transient spatial morphologies, including polygon-like patterns (Fig.~\ref{fig4}). Notably, these structures emerge spontaneously without externally imposed gradients, demonstrating that the intrinsic coupling between nonlinear reaction kinetics and diffusion is sufficient to drive spatial self-organization. Similar wavefront propagation phenomena have been reported experimentally in fuel-driven supramolecular systems~\cite{leira2018oscillations,kubota2020force}.

In contrast, for parameter values in the stable regime, the system remains spatially homogeneous and no pattern formation is observed even in presence of species dispersal. This demonstrates that spatiotemporal structure requires an underlying oscillatory instability. Although the diffusion coefficients depend on the local polymer state through length-dependent mobility, this coupling alone cannot generate patterns without the destabilizing effect of the reaction kinetics.
\begin{figure*}[t]
    \centering
    \includegraphics[width=\textwidth]{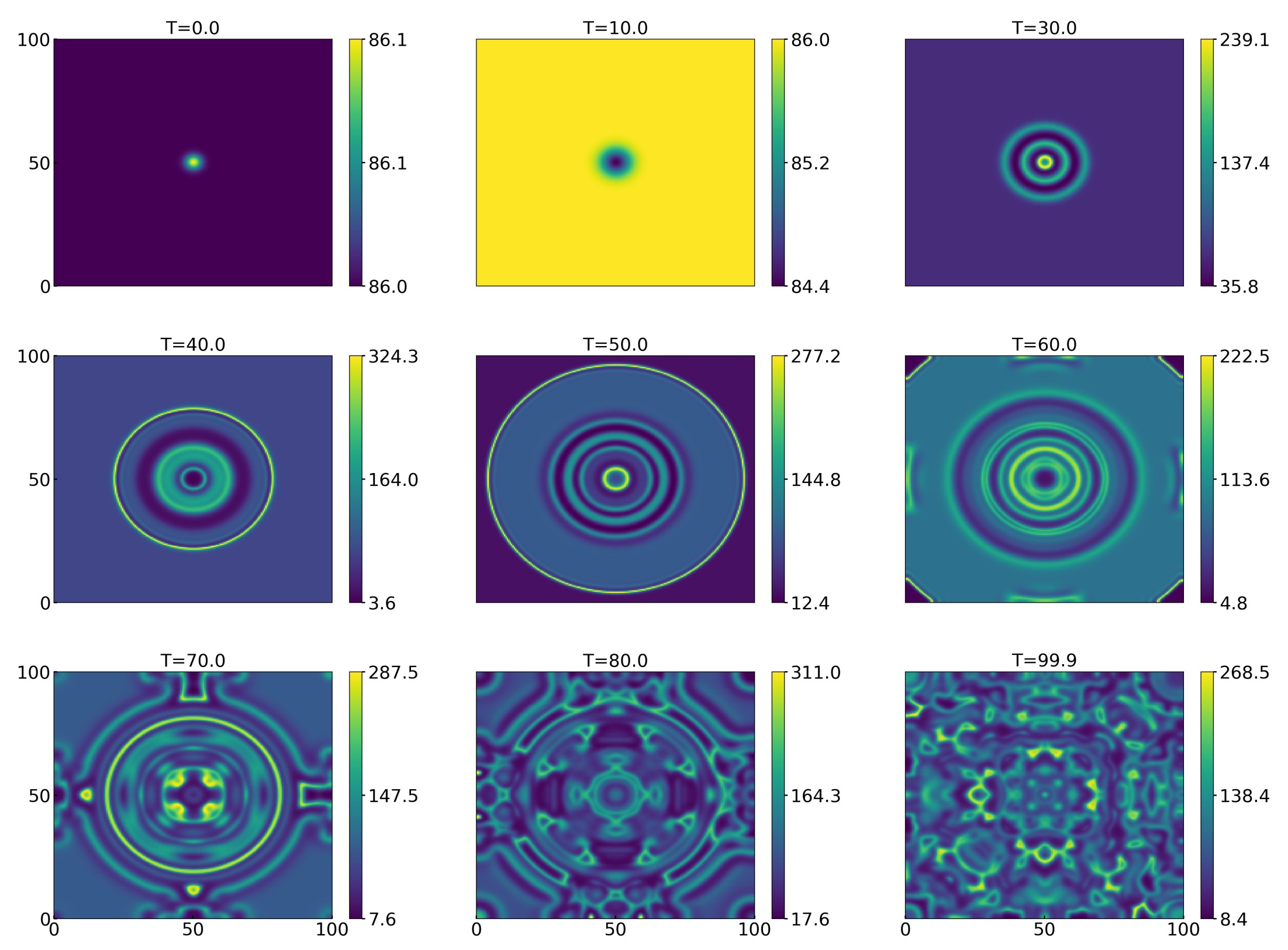}
\caption{
Wavefront dynamics initiated by a localized Gaussian perturbation, showing radially propagating fronts (polymer mass concentration) with internal structure and eventual transition to complex spatial patterns.
Parameters are taken from the oscillatory region: $k_{a}^\prime = 4.0$, $k_{d2}^\prime = 5000.0$, and $k_{d3}^\prime = 1.5$ and the domain size is $100 \times 100$ with a grid size of 0.1. Time is expressed in units of $(k_{+} K)^{-1}$.}
\label{fig5}
\end{figure*}

To further analyze wavefront dynamics, we introduce a localized Gaussian perturbation as a controlled initial condition. This generates radially symmetric wavefronts that propagate outward from the perturbation center (Fig.~\ref{fig5}), enabling a clearer characterization of front dynamics.
The propagating fronts exhibit intricate internal structures, including concentric regions of alternating high and low concentration that continuously form and dissipate during propagation. At longer times, the system again evolves into transient polygon-like patterns, consistent with those observed for random perturbations. The corresponding pattern evolution is further illustrated in Movie-S2 (see SI, Section-3). In addition, we explored multiple points across the parameter space and present the spatiotemporal dynamics of a representative parameter set ($k_{a}^\prime = 0.9$, $k_{d2}^\prime = 2000.0$, and $k_{d3}^\prime = 0.1$) located close to the bifurcation surface in Movie S3 (SI, Section-3). The qualitative nature of the wave dynamics closely resembles that observed deep within the spatiotemporally unstable domain.

These results demonstrate that wavefront propagation and pattern formation are robust with respect to initial conditions and parameter variations (see Figure S1 in SI for the variation of parameter $C$). Overall, the interplay between nonlinear reaction kinetics and diffusion of the polydisperse species provides a minimal and physically grounded mechanism for the emergence of self-organized wavefronts and complex spatiotemporal patterns in fuel-driven supramolecular systems.
To quantify this behavior and extract scaling properties, we next analyze the dynamics of the propagating fronts.

\subsection{Nonlinear propagation of polymerization fronts}

To quantitatively characterize the propagation dynamics of the traveling polymerization fronts, we extracted one-dimensional radial profiles of the polymer mass concentration \(m_{1}^\prime\) from the two-dimensional simulations at successive timesteps. Since the initial Gaussian perturbation is introduced at the center of the domain, the resulting fronts propagate approximately radially outward. The instantaneous front position \(R(t)\) was determined from the spatial derivatives of the concentration profile using a threshold-based procedure (see SI, Section~5 for details).

\begin{figure*}[h!]
    \centering
    \includegraphics[width=0.49\textwidth]{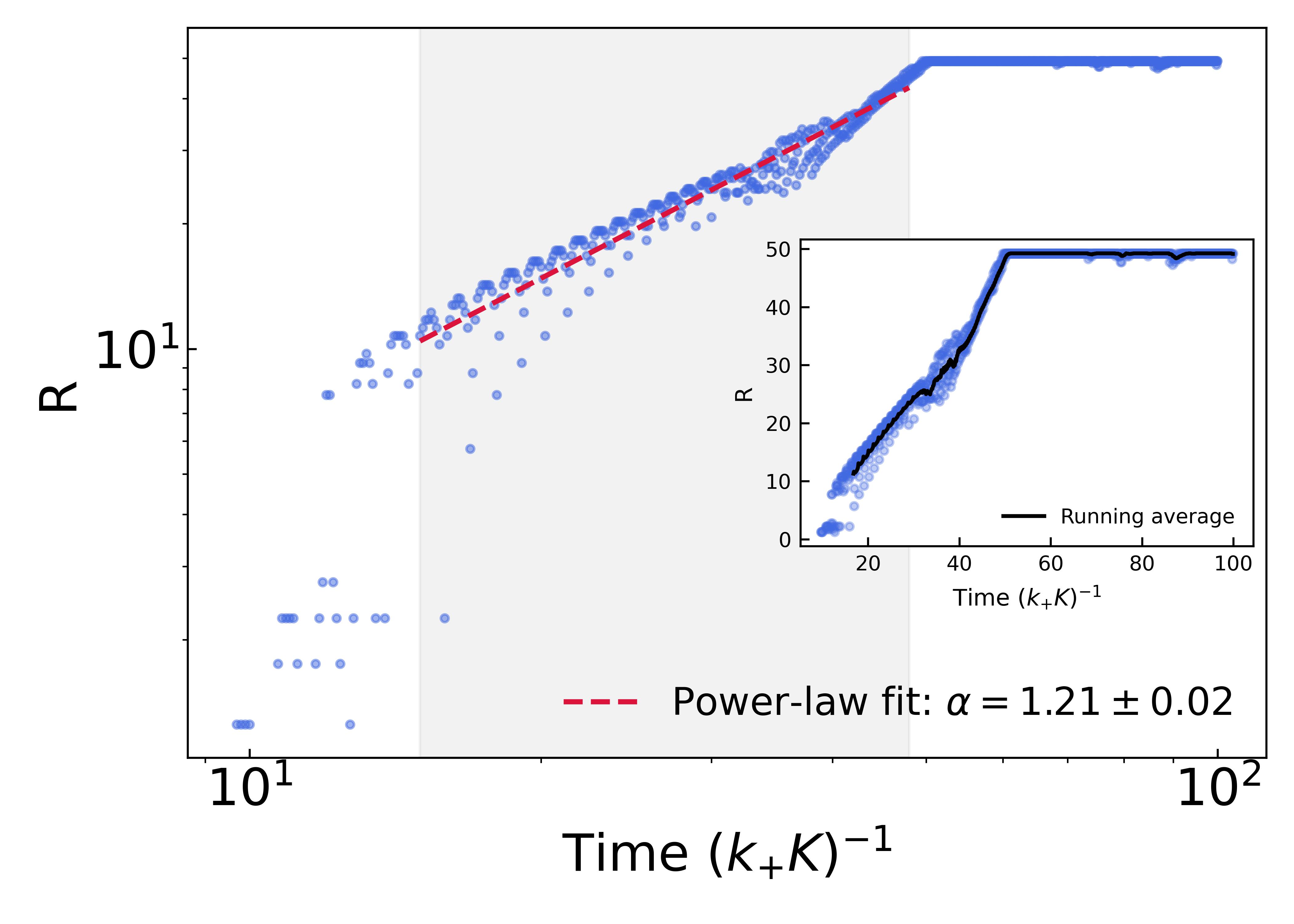}\hfill
     \includegraphics[width=0.49\textwidth]{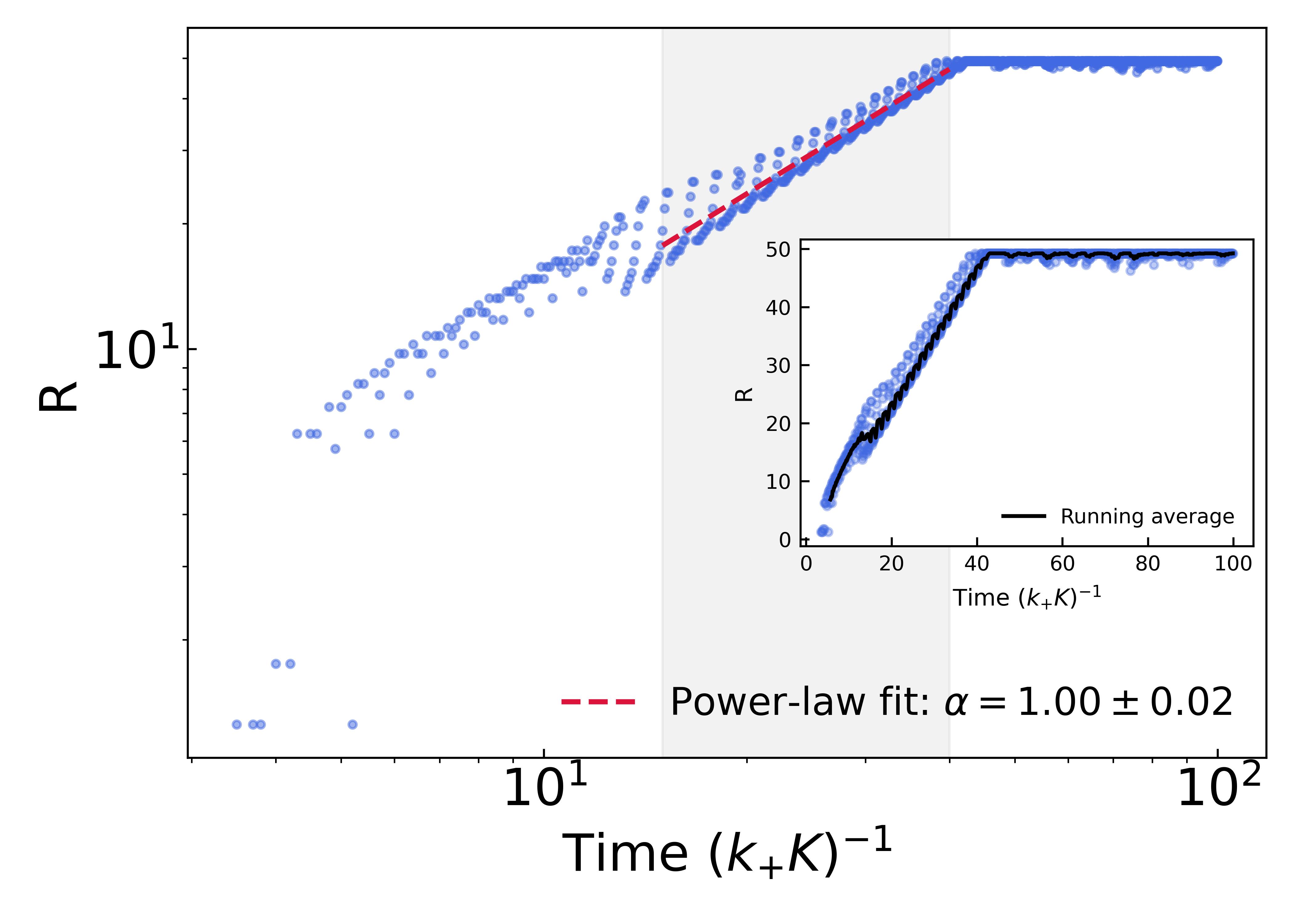}\quad
    \caption{
Temporal evolution of the propagating polymerization front for a Gaussian-perturbed system for (a) $k_{a}^\prime = 4.0$, $k_{d2}^\prime = 5000.0$, and $k_{d3}^\prime = 1.5$ and (b) near the bifurcation surface for $k_{a}^\prime = 0.9$, $k_{d2}^\prime = 2000.0$, and $k_{d3}^\prime = 0.1$ as shown on a log-log scale, indicating near-ballistic and ballistic spreading behavior respectively.
Inset: Representative radial profile of the polymer mass concentration used to determine the front position $R(t)$.
}
    \label{fig6}
\end{figure*}

The extracted front dynamics reveal nonlinear spreading characterized by the power-law relation $R(t)\sim t^{p}$, where the exponent \(p\) depends on the underlying kinetic parameters. For representative parameter sets, the propagation shows near-ballistic behavior (\(p\approx1\)) as shown in Fig.~\ref{fig6}. These dynamics are substantially faster than ordinary diffusive transport, for which \(R(t)\sim t^{1/2}\), indicating that front propagation is governed by the coupled interplay between reactive feedback and dynamically evolving transport.

A key feature of the model is the incorporation of polymer length-dependent diffusion through the relation \(D_m \sim 1/L_p\), where \(L_p=m_{1}^\prime/m_{0}^\prime\) is the local mean polymer length. Consequently, transport coevolves with the local assembly state. As polymer-rich regions form, the mobility of longer supramolecular chains decreases, slowing diffusive homogenization and thereby stabilizing propagating interfaces and local concentration gradients. In contrast, regions dominated by shorter polymers or monomers remain comparatively mobile, enabling rapid redistribution of material. The resulting feedback between supramolecular assembly and transport generates a dynamically heterogeneous medium in which wave propagation emerges self-consistently from the nonequilibrium polymerization dynamics.

\pg{To assess the role of state-dependent transport, we further compared the full length-dependent diffusion model with simulations employing constant polymer diffusivities, namely \(D_m=D_0\) and \(D_m=D_0/L_{ss}\), where \(L_{ss}\) is the steady-state mean polymer length. Qualitatively, all three cases exhibit traveling fronts and related spatiotemporal structures, indicating that the emergence of wave propagation is fundamentally reaction-driven. However, important quantitative differences arise in the propagation dynamics. In particular, simulations with \(D_m=D_0/L_{ss}\) closely reproduce the wavefront spreading (shown in Movie-S4 and S5 in SI) observed in the full state-dependent diffusion model, suggesting that the average reduction in polymer mobility already captures much of the effective transport behavior. By contrast, when \(D_m=D_0\), polymers remain highly mobile throughout the system, leading to faster front propagation and earlier interaction with the system boundaries. These comparisons demonstrate that while oscillatory reaction kinetics primarily drive pattern formation, polymer length-dependent diffusion plays a crucial role in quantitatively regulating front speed, gradient stability, and mesoscale transport dynamics.}

The observed nonlinear front propagation therefore arises from the combined interplay between activation--deactivation kinetics, cooperative growth, fragmentation, and state-dependent diffusion. As the polymer-rich front advances, local reaction feedback continuously regenerates the leading edge of the interface, sustaining coherent propagation across the system. Such self-organized front dynamics are characteristic of nonequilibrium reaction-transport systems in which chemical activity and transport are intrinsically coupled across space and time.

\subsection{Robustness to system size and numerical resolution}

To assess the robustness of the observed spatiotemporal dynamics, we examined the effects of system size and spatial resolution (see SI, Section~6). Increasing the domain size primarily affects boundary interactions, allowing a larger number of polymerization seeds to form and wavefronts to propagate over longer distances. However, the qualitative features of the dynamics including wavefront initiation, propagation, and pattern formation, remain unchanged. Representative results are shown in Figs.~S2 and S3 in SI.

We further verified that the observed behavior is not an artifact of spatial discretization. While coarse grids introduce minor anisotropies in the wavefront structure, finer grids and higher-order discretization schemes recover isotropic propagation and preserve the same qualitative dynamics (Figs.~S4--S5).

These results confirm that the emergence of wavefronts and spatiotemporal patterns is an intrinsic property of the model and robust across system size and numerical resolution.

\subsection{\pg{Experimental implications and design principles}}

\pg{The present reaction--transport model captures key dynamical behaviors observed in chemically fueled supramolecular polymerization experiments, including autonomous oscillations, traveling fronts, and large-scale spatiotemporal organization~\cite{leira2018oscillations,kubota2020force}. In both experiments and the present theoretical framework, these behaviors emerge from the interplay between autocatalytic growth and delayed deactivation, consistent with the general requirements for self-sustained oscillations in driven chemical systems~\cite{epstein1998introduction,nicolis1977self}. However, the model also predicts several experimentally accessible behaviors that extend beyond those reported so far.}

\par \pg{\textit{Chemical wavefronts in convection-suppressed environments:} In existing supramolecular polymerization experiments, centimeter-scale patterns are often strongly influenced by density-driven convection~\cite{leira2018oscillations}. In contrast, the present model generates propagating fronts and structured spatiotemporal patterns purely through reaction--transport coupling in the absence of flow. In particular, the simulations predict radially propagating multi-ring wavefronts (Fig.~\ref{fig5}) arising from the interplay between oscillatory kinetics and polymer length-dependent diffusion. We therefore predict that similar fuel-driven supramolecular polymerization system, when studied under convection-suppressed conditions such as thin gels, confined Hele--Shaw geometries, or microfluidic environments, should exhibit qualitatively distinct diffusion-dominated wave structures. Observation of such concentric fronts would provide a direct experimental test of the reaction--transport mechanism proposed here.}

\par \pg{\textit{Kinetic control of oscillatory and transport regimes:} The model further provides a quantitative phase diagram identifying the transition between steady and oscillatory regimes through a Hopf bifurcation. This predicts that experimentally tunable parameters, such as fuel concentration, activation rate, or fragmentation/deactivation kinetics, can be used to switch oscillatory dynamics on or off in a controlled manner. In addition, the propagation dynamics of the polymerization fronts depend sensitively on the kinetic regime: systems close to the bifurcation exhibit near-ballistic front propagation, whereas deeper oscillatory regimes display faster spreading dynamics associated with stronger nonlinear feedback. These results suggest that macroscopic transport behavior can be systematically regulated through molecular-scale kinetic design.}

\par \pg{More broadly, the present framework provides a controllable and computationally accessible platform for exploring how nonequilibrium supramolecular assembly generates emergent spatiotemporal behavior. Beyond reproducing experimentally observed oscillations and fronts, the model predicts structured wave dynamics and transport regimes that remain experimentally unexplored. The compactness of the framework: four dynamic variables, three principal kinetic parameters, and a single physically motivated transport ingredient, makes these predictions both mechanistically transparent and experimentally testable. Overall, the model establishes minimal design principles for engineering programmable supramolecular materials capable of autonomous oscillation, wave propagation, and spatiotemporal self-organization.}

\section{Conclusions and Outlook}

The emergence of self-organized patterns in dissipative supramolecular systems has become a central theme in systems chemistry, offering a route to life-like spatiotemporal organization~\cite{walther2020responsive,leira2018oscillations}. In this work, we introduced a minimal reaction--diffusion model for a fuel-driven supramolecular polymerization system that captures the interplay between activation--deactivation chemistry and cooperative assembly--disassembly dynamics. Despite its simplicity, the model reproduces a broad range of experimentally relevant behaviors, including autonomous oscillations, propagating wavefronts, and complex transient spatial patterns.

A major challenge in modeling such systems lies in coupling nonlinear reaction kinetics with the transport of monomers and dynamically evolving polymeric species. Our framework addresses this by incorporating polymer length-dependent diffusion based on Rouse scaling, providing a physically motivated description of mobility in polydisperse supramolecular assemblies. Linear stability analysis identifies a Hopf bifurcation separating steady and oscillatory regimes, while the inclusion of diffusion extends these temporal instabilities into rich spatiotemporal dynamics characterized by traveling fronts and transient polygonal and wave-like structures.

Importantly, the model enables quantitative characterization of front propagation and reveals propagation dynamics follows a near-ballistic spreading. This behavior emerge from the nonlinear coupling between oscillatory reaction kinetics and state-dependent transport, where local supramolecular assembly dynamically modulates mobility. The resulting reaction--transport feedback establishes a direct connection between molecular-scale assembly processes and emergent mesoscale dynamics.

Several features predicted by the model may also be experimentally accessible. In particular, the predicted concentric multi-ring wavefronts could potentially be explored in environments where convective flow is suppressed, such as gels or thin microfluidic films. More broadly, the framework provides qualitative design principles for programming oscillatory and propagating behavior in fuel-driven supramolecular materials through control of activation, fragmentation, and transport processes.

\pg{Overall, this work demonstrates that minimal nonequilibrium reaction networks, when coupled to physically motivated transport, are sufficient to generate complex spatiotemporal organization reminiscent of biological systems. By identifying the mechanistic routes through which simple chemical processes produce oscillations, propagating fronts, and pattern formation, the model contributes to a growing theoretical foundation for programmable dissipative self-assembly and active supramolecular matter.}

\pg{Future extensions may incorporate additional physical effects such as hydrodynamic coupling, external fields, or stochastic fluctuations to further connect reaction kinetics with transport and thermodynamic irreversibility. Experimental validation of the predicted structured wave dynamics and transport regimes would provide important tests of the framework and inspire new experimental strategies for engineering life-like supramolecular materials capable of autonomous spatiotemporal organization.}

%

\section*{Conflicts of interest}
The authors have no conflicts to declare.

\section*{Code Availability}
The main simulation code developed to generate the results reported in this study are available at \url{https://github.com/JMLab-tifrh/supramolecular_polymer_patterns}.

\section*{Supporting Information}
The Supporting Information (SI) provides additional figures (Figs. S1-{S6}) and supporting movies S1-S5 mentioned in the main text, along with supplementary methods and technical details that support the findings of this manuscript. 

\section*{Acknowledgements}
The authors acknowledge the Tata Institute of Fundamental Research Hyderabad, and the Indian Institute of Science Education and Research Thiruvananthapuram, India, for providing computing resources. The authors acknowledge support from the Department of Atomic Energy, Government of India, under Project Identification No. RTI 4007. A.S. thanks Hridey Narula, Roshan Maharana, and Purnima Jain for useful discussions.

\bibliography{ref}

\end{document}